\documentclass[doublecol]{epl2}

\usepackage{amssymb}
\usepackage{amsmath}

\newcommand{\tkone}{T_K^{S=1}}

\newcommand{\ve}{\varepsilon}

\newcommand{\bea}{\begin{eqnarray}}
\newcommand{\eea}{\end{eqnarray}}

\def\beq{\begin{equation}}
\def\eeq{\end{equation}}
\newcommand{\ket}[1]{| #1 \rangle}

\begin{document}

\title{Quantum Transport Through a Stretched Spin--1 Molecule}

\author{P. S. Cornaglia\inst{1}, P. Roura Bas\inst{2}, A. A. Aligia\inst{1}, C. A. Balseiro\inst{1}}
\shortauthor{P. S. Cornaglia, P. Roura Bas, A. A. Aligia, C. A. Balseiro}

\institute{                    
	\inst{1} Centro At\'omico Bariloche and Instituto Balseiro, CNEA, 8400 Bariloche, Argentina and\\	Consejo Nacional de Investigaciones Cient\'{\i}ficas y T\'ecnicas (CONICET), Argentina\\
	\inst{2}Centro At\'omico Constituyentes, Comisi\'on Nacional de Energía At\'omica, Buenos Aires, Argentina
}

%\date{\today}

\abstract{
We analyze the electronic transport through a model spin--1 molecule as a function of temperature, magnetic field and bias voltage. 
We consider the effect of magnetic anisotropy, which can be generated experimentally by stretching the molecule.
In the experimentally relevant regime the conductance of the unstretched molecule reaches the unitary limit of the underscreened spin--1 Kondo effect at low temperatures. 
The magnetic anisotropy generates an antiferromagnetic coupling between the remaining spin 1/2 and a singular density of quasiparticles, producing a second Kondo effect and a reduced conductance.
The results explain recent measurements in spin--1 molecules [Science {\bf 328} 1370 (2010)].
}
%73.23.-b	Electronic transport in mesoscopic systems
%72.15.Qm	Scattering mechanisms and Kondo effect
\maketitle
Recent experimental advancements in measurement and control of molecular devices open the possibility of studying exotic electronic behavior in a controlled way and allow for a detailed comparison with the predictions of strongly correlated electron theories.
In molecular junction experiments, a single molecule is contacted to two metallic source and drain electrodes and its magnetic and electronic properties can be controlled using a capacitively coupled gate electrode~\cite{Park2002}, applying an external magnetic field or mechanically by stretching the molecule~\cite{Parks06112010}.  
The reduced size of the molecules leads to strong electron-electron interactions, which give rise to Coulomb blockade effects, and strongly correlated electron phenomena as the spin--1/2 ~\cite{Park2002} and the underscreened spin--1 Kondo effects~\cite{Roch2008,PhysRevLett.103.197202,Parks06112010}. 

The recent observations of the underscreened spin--1 Kondo effect~\cite{Roch2008,PhysRevLett.103.197202,Parks06112010} are particularly interesting because the ground state of the system is not a Fermi liquid, as in the fully screened Kondo effect~\cite{Nozieres1974}, but a singular Fermi liquid composed by a diverging density of magnetic excitations at low energy, together with an asymptotically free spin 1/2~\cite{PhysRevB.75.245329,PhysRevB.72.014430,Hewson2005Sing}. 
In addition, in these experiments the ground state of the system can be modified and a quantum phase transition induced applying external fields.

In this paper we model the experiments by Parks {\it et al.} \cite{Parks06112010} in which a spin--1 molecule, in the underscreened Kondo regime, was stretched while measuring the electronic transport through it. 
We first show that an axial stretching of the molecule leads to a magnetic anisotropy term in the Hamiltonian which changes dramatically the low temperature transport properties. The magnetic anisotropy couples the remaining unscreened spin 1/2 and the local singular Fermi liquid excitations. As a consequence, the molecule shows a two stage Kondo effect  to a Fermi liquid ground state. Stretching the molecule drives a Kosterlitz-Thouless quantum phase transition from a high-conductance singular Fermi liquid to a low-conductance Fermi liquid ground state. 
%ACA
Applying an external magnetic field parallel to the anisotropy axis, a crossing of the lowest lying molecular levels is induced and the conductance increases again. 

%ACA
In the Co(tpy-SH)$_2$ complex studied in Ref. \cite{Parks06112010}, the Co atom is in the center of a nearly perfect octahedron of six N atoms.
%the Co atom is subject to an environment of approximately octahedral symmetry. 
This splits the d levels of Co into three lower-energy $t_{2g}$ and two higher-energy $e_g$ orbitals. 
To analyze the effect of stretching the molecule, we have considered 
a local model Hamiltonian which contains all interactions inside
the d shell plus octahedral $H_{O}$ and tetragonal $H_{T}$ crystal
fields (as described in Ref. \cite{kroll}), and we have included the
spin-orbit interaction $\lambda H_{SO}$ in second-order perturbation theory.

In absence of $H_{SO}$, the ground state of the d$^{8}$ configuration of 
Co$^{1+}$ is a $B_{1g}$ triplet with a hole in each $e_g$ orbital. 
%(in terms of $e_{g}$ holes, $|11\rangle
%=d_{x^{2}-y^{2},\uparrow }^{\dagger }d_{3z^{2}-r^{2},\uparrow }^{\dagger }$
%and its SU(2) partners). 
The spin-orbit coupling mixes this state with several excited
singlet and triplets which contain one $e_{g}$ and one $t_{2g}$ hole. As a
consequence, the $S_z=0$ projection state of the triplet $|T,0\rangle $  
is split from the non-zero projection states $|T,\pm 1\rangle $.
The energy difference is $D=\lambda ^{2}f$, where $f$ depends mainly on $H_{T}$ 
and the Coulomb integrals which characterize the
interactions inside the d shell \cite{kroll}. 
This splitting can be described by a $DS_{z}^{2}$ term in the Hamiltonian. 
The tetragonal field $H_{T}$ affects mainly the  $e_{g}$ orbitals, 
which point towards the six ligand atoms.
When the octahedron of these ligand atoms is elongated in the $z$ direction (as in La$_{2}$CuO$_{4}$)
it is more favorable energetically to put the holes in the $d_{x^{2}-y^{2}}$
orbitals than in the $d_{3z^{2}-r^{2}}$, leading to positive 
$D$ ($|T,0\rangle $ is favored), while when the octahedron is compressed in the
same direction, the situation is the opposite, as in the Haldane system 
Y$_{2}$BaNiO$_{5}$, where $D<0$ \cite{payen}.

Taking the values of the Coulomb integrals and $\lambda $ 
which fit the low energy spectra of the neutral Ni atom 
(which has a d$^8$ electronic configuration as Co$^{+1}$), an $H_O$ splitting of $2$ eV, and a
splitting of the $e_{g}$ levels of 1 eV (favoring $d_{x^{2}-y^{2}}$ holes),
we obtain $D$ $\sim 0.83$ meV, a value which is consistent with the
typical experimental observations.

To study the electronic transport through the magnetic molecule we consider the effective Hamiltonian
$H=H_M+H_E+H_{V}$,
where
\begin{eqnarray}
H_M &=& \sum_{\ell=a,b} \left[U_\ell n_{\ell \uparrow} n_{\ell%
\downarrow}+ \varepsilon_\ell (n_{\ell \uparrow} +n_{\ell%
\downarrow})\right] \nonumber\\ &+& J \mathbf{S}_a \cdot \mathbf{S}_b%
 -\mu_B {\mathbf H \cdot \mathbf S} + DS_z^2,
\end{eqnarray}
describes the two effective e$_g$ levels $(a, b)$ of the molecule, which are relevant for the electronic transport, coupled through a Hund rule ferromagnetic exchange $J<0$,  with a stretching induced anisotropy $D$. 
We will focus on the parameter regime where the molecular ground state is in the spin $S=1$ sector: $|T,S_z\rangle$ with energy $E_T=2 \varepsilon +J/4 + S_z(D S_z- \mu_B H)$.

The Hamiltonian of two non-interacting source and drain leads is given by 
$H_{E}=\sum_{{\bf k},\sigma} \varepsilon_{\alpha}({\bf k})\; c^{\dagger}_{{\bf k} \sigma \alpha} c^{}_{{\bf k} \sigma \alpha}$, with $\alpha=L,R$,
and the coupling between the molecule and the leads is described by the last
term in the Hamiltonian, 
\begin{equation}
H_{V}=\sum_{{\bf k\ell},\sigma,\alpha}\;V_{{\bf k}\alpha\ell}\left(d^{\dagger}_{\ell\sigma}\;c^{}_{{\bf k}\sigma\alpha} + c^{\dagger}_{{\bf k}\sigma\alpha}\;d^{}_{\ell\sigma} \right)\;. \nonumber
\end{equation}
In order to model the experimental observations, we will consider that a single screening channel is relevant. 
A second channel would lead to a complete screening of the molecular spin, although in general at exponentially small temperatures~\cite{PhysRevB.75.245329}. 
We assume that only one of the molecular levels is coupled to the electrodes, and in what follows we take $V_{{\bf k}R b}=V_{{\bf k}L b}=0$. This can be done without loss of generality in the limit of large $U$, since other configurations are related by a level rotation.

For symmetric hybridization to the electrodes, the conductance through the system is~\cite{PhysRevB.50.5528}
\begin{equation}\label{eq:cond}
G(T)=\frac{e^{2}}{h} \Delta \pi
\sum_{\sigma}\int_{-\infty}^{\infty} d\omega \left(-\frac{\partial
f(\omega )}{\partial \omega }\right)\rho^{\sigma}_{aa}(\omega ),
\end{equation}
where $\rho^{\sigma}_{aa}(\omega)$ is the local electronic density
of states on level $a$.  Here, $\Delta= 2 \pi \rho_0\langle V_{\bf
k}^2\rangle$ where $\rho_0$ is the electronic density of states per spin of
the electrodes at the Fermi level, the brackets denote the average
over the Fermi surface, and $f(\omega)$ is the Fermi function.

When the anisotropy term is positive $D>0$ the ground state of the 
system is a Fermi liquid~\cite{Hewson2005Sing,PhysRevB.78.224404} and 
we can obtain an exact expression for zero temperature
$a$-level spectral density for spin $\sigma$ in the wide band limit~\cite{PhysRevB.71.075305},
\begin{eqnarray}
\rho^\sigma_{aa}(0)= \frac{1}{\pi \Delta}  \sin^2 \left[\pi\left(n_{a}^\sigma + n_{b}^\sigma \right)\right]\;,
\label{rhof}
\end{eqnarray}
%where $n_{\sigma}=n_a+n_b$ is the charge in the levels $a$ and $b$.
where $n_{\ell}^\sigma$ is the charge in the level $\ell$ with spin $\sigma$.
Equations~(\ref{eq:cond}) and (\ref{rhof}) lead to the following
expression for the zero temperature conductance~\cite{Glazman2001,PustilnikGH2003,PhysRevB.71.075305}:
\begin{eqnarray}\label{eq:condFL}
g\equiv\frac{G}{G_0} =\frac{1}{2} \sum_{\sigma}\;\sin^2\left[\pi
\left(n_{a}^\sigma + n_{b}^\sigma \right)\right]\;,
\end{eqnarray}
where $G_0=2 e^2/h$ is the quantum of conductance.

In absence of magnetic field
$n_\ell^\sigma=n_\ell^{-\sigma}$ and $g=\sin^2\left[\frac{\pi}{2}
\left(n_{a} + n_{b} \right)\right]$ where $n_{\ell} = \sum_\sigma
n_{\ell}^\sigma$. The zero-temperature conductance thus vanishes
when the total number of electrons in the two levels of the molecule is even.
From now on we will focus on charge sector where $n_a+n_b\simeq 2$, so that we expect a small conductance at low temperatures whenever the ground state is a Fermi liquid. 
In presence of a magnetic field $H$, the molecule will have a
 magnetization $m=\frac{1}{2} (n_{\uparrow}-n_{\downarrow})$
 and  Eq.~(\ref{eq:condFL}) becomes
\begin{equation}
g = \sin^2(\pi m).
\label{eq:luttinger-m}
\end{equation}

\begin{figure}[t]
  \includegraphics[width=0.9\linewidth,clip=true]{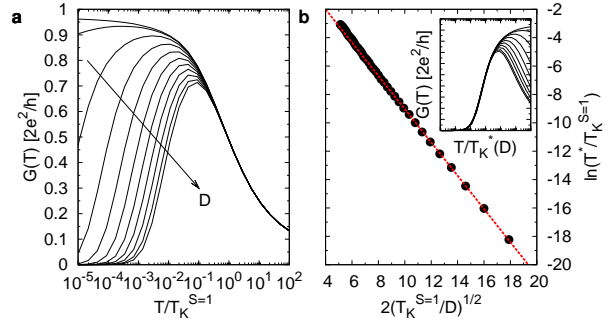}
  \caption{a) Zero-bias conductance through the molecule vs. temperature for different values of the magnetic anisotropy: 
	$D/\tkone= 0.156,\ldots, 0$, with $\delta D/\tkone=0.0156$ between consecutive curves. Other parameters are: $U=0.25W$, $\varepsilon=0.125W$, $\Delta=0.035W$, and $J=-0.005W$ with $W$ the conduction electron bandwidth and the unit of energy. b) Kondo temperature $T_K^\star$. Fit using Eq. (\ref{eq:tstar}), the fitting parameters are $c_1\sim 0.77$, $c_2\sim 1.07$. Inset: Same as (a) with each curve scaled by $T_K^\star$.\label{fig:tks}}
\end{figure}

At finite temperatures we have calculated the zero-bias conductance using the numerical renormalization group (NRG)~\cite{RevModPhys.80.395}\footnote{For the NRG calculations we kept up to 1600 states at each iteration and used a energy discretization parameter $\Lambda=2.5$.}.  
Figure \ref{fig:tks}(a) shows the zero-bias conductance as a function of temperature
for several values of the anisotropy and $H=0$.
In absence of anisotropy,
the molecular spin is partially screened by the conduction electrons below a Kondo temperature $T_K^{S=1}$.
The development of the underscreened Kondo effect is associated with a monotonic increase in the conductance which, for $D=0$, reaches the unitary limit at zero temperature. 
This is in stark contrast to what is expected for a Fermi liquid from Eq. (\ref{eq:condFL}). In fact, the ground state of the system is a singular Fermi-liquid~\cite{PhysRevB.72.014430}.
The conductance, for $D=0$, is an universal function of $T/\tkone$: $G(T)=G_0f(T/\tkone)$ that differs from the one obtained for a fully screened $S=1/2$ Kondo effect \cite{Hewson2005Sing}.

When a positive anisotropy ($D>0$) is turned on, 
the ground state of the isolated molecule is the $S_z=0$ state $\ket{T,0}$. 
In this case we expect a Fermi liquid ground state~\cite{PhysRevB.78.224404} for the system, and a small conductance at low temperatures.
For $D\ll k_B\tkone$ we observe first an increase in the conductance as the temperature is lowered, followed by a plateau of high conductance and a reduction of the conductance for temperatures of the order of a characteristic temperature $T_K^\star\ll D/k_B$, where $T_K^\star$ is defined by $G(T/T_K^\star)=0.5$. 
We note that the low-temperature region of the $G(T)$ curves collapses into a single universal curve when the temperature is scaled by $T_K^\star(D)$ [see inset in Fig.~\ref{fig:tks}(b)]. 
In the regime where $D> k_B\tkone$, the decrease in the conductance occurs at temperatures $T\sim D/k_B$, where the energy gap $D$ between the $S_Z=\pm 1$ and the lowest lying $S_z=0$ dominates the physics and cuts-off the underscreened Kondo effect.

From the plot of $\ln(T_K^\star)$ as a function of $(\tkone/D)^{1/2}$ [see Fig. \ref{fig:tks}(b)] 
it is seen immediately that a good fit of $T_K^\star$ can be obtained using the formula
\beq\label{eq:tstar}
T_K^\star=c_1 \tkone e^{-c_2 2\sqrt{\frac{\tkone}{D}}},
\eeq
where in practice $c_1,c_2\sim 1$ and depend weakly on the model parameters. 
This temperature scale can be identified with the Kondo temperature 
of a second stage Kondo effect, induced by the magnetic anisotropy, 
in which the remaining spin 1/2 is screened. For $D\to0$ there is quantum phase transition of the Kosterlitz-Thouless type from Fermi liquid to singular Fermi liquid. 
%ACA
%and to non-Fermi liquid for negative $D$. 
A similar behavior is obtained for the singlet-triplet quantum-phase transition in  models of magnetic impurities \cite{allub,VojtaQPTimp,roura2} and quantum dots~\cite{Glazman2001,PustilnikGH2003,HofstetterQPT,Koganexp2003,logan}. 
Note, however, that the singlet state (ruled out by ab initio calculations) and a second screening channel (ruled out in Ref. \cite{Parks06112010} by experiments with an external magnetic field at different angles with the stretching axis) are absent in our model and the low-temperature screening is of a different nature.

To analyze the effect of the anisotropy term, we will focus in the regime of large ferromagnetic $|J|$. In this regime the spin of the molecule is one, and we can describe it as a sum of two spin--1/2: $\mathbf{S}_1$ and $\mathbf{S}_2$. Using $\mathbf{S}^2=(\mathbf{S}_1+\mathbf{S}_2)^2=2$ we can rewrite the anisotropy term as
\beq\label{eq:anti}
H_D=DS_z^2=3D/4 +D(S_{1z}S_{2z}-S_{1x}S_{2x}-S_{1y}S_{2y}),
\eeq
which describes an anisotropic Kondo coupling. For $D=0$, we can consider that one of the two spins is fully screened at temperatures $ T \ll \tkone$ and the other is asymptotically free \cite{VojtaQPTimp}. When an anisotropy term $D\ll k_B\tkone$ is turned on, the spin will be coupled through Eq. (\ref{eq:anti}) to the singular Fermi liquid that results from the underscreened Kondo effect. $D$ flows to strong coupling at low energies leading to a second stage Kondo effect. 
This situation is analogous to the one obtained with the same model, but with a positive exchange coupling $J$, where a two stage Kondo effect is obtained~\cite{PhysRevB.71.075305}. In that case the remaining spin couples antiferromagnetically to a local Fermi liquid that results from the fully screened spin-1/2 Kondo effect and has a quasiparticle density of states:
$\rho^s_{QP}=\frac{T_K/\pi}{\omega^2+T_K^2}$,
where $T_K$ is the Kondo temperature of the first Kondo stage.
Performing a slave boson mean field approximation (SBMFT) it is then possible to obtain the Kondo temperature for the second stage Kondo effect using the formula~\cite{PhysRevB.71.075305}:
\beq \label{eq:TK}
\frac{2}{J}=\int_{-\infty}^{\infty}\frac{d\ve}{\ve}\tanh(\ve/2T_0)\rho_{QP}^{s}(\ve)\,.
\eeq
The resulting second stage Kondo temperature is given by $T_0\sim T_Ke^{-\pi T_K/J}$ and has an excellent agreement with the numerical results~\cite{PhysRevB.71.075305}. 
In the present case, when the anisotropy term is present, the ground state of the system is a Fermi liquid and we expect the standard SBMFT approximation to give a good description of the low energy physics. 
Following the same reasoning we can extract $T_K^\star$ making the replacement: $T_0\to T_K^\star$, $J\to D$ and $\rho_{QP}^{s}\to\rho^{us}_{QP}(\ve)$ in Eq.~(\ref{eq:TK}), where $\rho^{us}_{QP}$ is the density of quasiparticles resulting from the underscreened Kondo effect, to which the spin--1/2 is antiferromagnetically coupled. To recover the expression of Eq.~(\ref{eq:tstar}) we need to assume the following density of quasiparticles which has an integrable divergence at low energy
\begin{equation}
\rho^{us}_{QP}(\ve)=\left\{\begin{matrix}\frac{-1}{2\tkone}\ln(|\ve|/\tkone) & \text{if}\;\;|\ve|< \tkone\\ 0&\text{if}\;\;|\ve|\geq \tkone\end{matrix}\right..
\end{equation}
\begin{figure}[!t]
  \includegraphics[width=0.9\linewidth,clip=true]{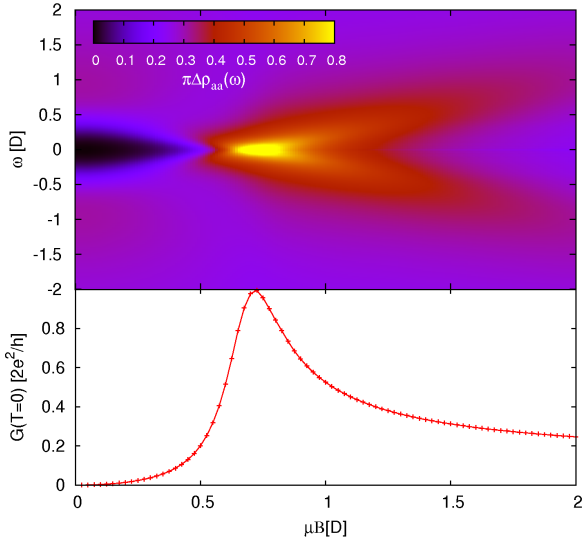}
  \caption{Spectral density $\rho_{aa}(\omega)$ (top) and zero-bias conductance (bottom) as a function of 
magnetic field for $D\sim 3 T_K$. Other parameters as in Fig. \ref{fig:tks}. \label{fig:mapBA}}
\end{figure}

%ACA 
%Another way to obtain a $D^{-1/2}$ in the exponent of Eq. (\ref{eq:tstar}) is to perform a poor man's scaling \cite{ander} of the Kondo model, including the anisotropy given by Eq. (\ref{eq:anti})  in the limit of $D/J\to 0$

%A non-monotonous behavior of the conductance as a function of $T$ is observed experimentally in Ref.~\cite{Parks06112010} for the stretched molecule. 
Figure \ref{fig:mapBA} shows a map of the spectral density $\rho_{aa}(\omega)$ as a function of energy and magnetic field. At $H=0$ we observe a splitting of the spectral density of order $D$, since in this case $D\sim 3k_B\tkone$. For $D\ll k_B\tkone$, the splitting is of order $T_K^\star$. As the magnetic field is increased, the splitting is reduced and the conductance increases. 
The zero temperature conductance is proportional to $\rho_{aa}(0)$ and can be analyzed using the Fermi liquid relation. When a magnetic field is applied the molecule starts to polarize and the conductance increases. In the low field limit $g$ increases quadratically with $H$,
\begin{equation}
g \approx \pi^2 m^2 \approx \pi^2 H^2 \chi^2
\label{g-de-B}
\end{equation}
were $\chi$ is the spin susceptibility with $\chi\propto 1/D$ for $D\gtrsim k_B\tkone$, and $\chi \propto 1/T_K^\star$for $D\ll k_B\tkone$.

For a magnetic field such that $\mu_BH \sim D$ the $\ket{T,1}$ and $\ket{T,0}$ molecular levels can be tuned to be degenerate. In this case an orbital Kondo effect takes place and the conductance reaches the unitary limit $g=1$ as expected for $m=1/2$, and the spectral density shows a Kondo peak at $\omega=0$.  
For very large $H$ the molecule becomes fully polarized, $m \to 1$, and the conductance also vanishes.

\begin{figure}[!t]
  \includegraphics[width=0.9\linewidth,clip=true]{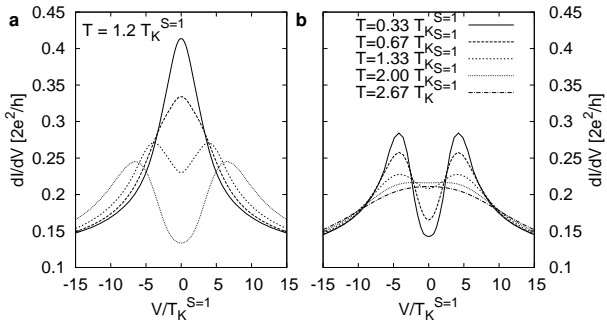}
  \caption{a) Differential conductance versus bias voltage for different values $D/k_B\tkone\simeq 0,\;0.36,\;0.6,\;0.96$ (from top to bottom at $V=0$). Other parameters are $\ve=0$, $J=0.6W$, and $2\Delta=0.141W$  b) Idem a) for $D\simeq 3.6 T_K^{S=1}$ and different temperatures. \label{fig:nca}}
\end{figure}

Finally, we turn our analysis to the out-of-equilibrium differential conductance at finite temperature. 
In the limit of large $U$ and $|J|$, only 5 molecular states are relevant for the transport properties, the three projections of the spin 1 and the two projections of a spin doublet with the electron localized at level $b$.
%ACA
This model was proposed for Tm impurities \cite{PhysRevLett.47.274} and solved exactly \cite{PhysRevB.33.6476}.
We used the non-crossing approximation \cite{roura2} to calculate the conductance
out of equilibrium. 
Figure \ref{fig:nca} shows the differential conductance as a function of bias voltage for different anisotropies as a function of the magnetic anisotropy at finite temperature (left panel), and for a fixed anisotropy term as a function of the temperature (right panel). As expected from the results of the zero temperature spectral density, the zero bias anomaly peak (ZBA) is split by the anisotropy term at low temperatures. At intermediate temperatures however it produces a reduction and a broadening of the ZBA.
The results are in good qualitative agreement with the experimental curves as a function of stretching and temperature~\cite{Parks06112010}.

In summary we have constructed a model to study the transport through a mechanically stretched magnetic molecule. We have shown that the stretching leads to a magnetic anisotropy term in the Hamiltonian which changes the low energy electronic properties of the molecular junction. The model reproduces qualitatively the experimental behavior of the transport properties as a function of temperature, bias voltage and magnetic field. This system allows for a detailed study of non-conventional electron behavior and magnetism in strongly correlated systems.

This work was partially financed by PIP No. 11220080101821 CONICET, and PICT R1776 of the ANPCyT.

\bibliography{S1}

\end{document}